\documentclass[aps,prb,superscriptaddress,preprintnumbers,
showpacs,legalpaper,twoside,twocolumn]{revtex4}

\usepackage{bm}
\usepackage{graphicx}
\usepackage{amsfonts}
\usepackage{amsmath}
\usepackage{amssymb}

\begin{document}

\title{Field-induced quantum disordered phases in $S=1/2$ 
weakly-coupled dimer systems with site dilution}

\author{Tommaso Roscilde}
\affiliation{Max-Planck-Institut f\"ur Quantenoptik, Hans-Kopfermann-strasse 1,
85748 Garching, Germany}
\affiliation{Department of Physics and Astronomy, University of Southern
California, Los Angeles, CA 90089-0484}


\begin{abstract}
In the present paper we discuss the rich phase diagram of 
$S=1/2$ weakly coupled dimer systems with site dilution 
as a function of an applied uniform magnetic field. 
Performing quantum Monte Carlo simulations on a 
site-diluted bilayer
system, we find a sequence of three distinct \emph{quantum-disordered} 
phases induced by the field. Such phases divide a doping-induced  
order-by-disorder phase at low fields from 
a field-induced ordered
phase at intermediate fields. The three quantum disordered
phases are: a gapless \emph{disordered-free-moment} phase, 
a gapped \emph{plateau} phase,
and two gapless \emph{Bose-glass} phases. 
Each of the quantum-disordered 
phases have distinct experimental signatures that make them 
observable through magnetometry and neutron scattering 
measurements. In particular the Bose-glass phase is characterized
by an unconventional magnetization curve whose field-dependence is 
\emph{exponential}. Making use of a local-gap model, we directly relate 
this behavior to the statistics of rare events in the system.  
\end{abstract}
\maketitle

 The $T=0$ field-induced ordering transition in spin-gap antiferromagnets
is one of the most intensively studied quantum phase transitions in
condensed matter systems, both theoretically and experimentally
\cite{Rice02}.
Examples of field-induced ordering can be found in Haldane chains
\cite{Zheludev05,Hondaetal}, 
and unfrustrated $S=1/2$ weakly-coupled dimer systems 
arranged in spin ladders \cite{Chaboussantetal97,Watsonetal01}, 
in coupled bilayers \cite{Jaimeetal04,Sebastianetal05}, 
and in more complex 3$d$ geometries \cite{Nikunietal00,Rueggetal03}.
 The application of a uniform
field overcoming the spin gap brings these systems from a gapped $S=0$ state 
to a state with finite magnetization parallel to the field 
and (in $d>1$) spontaneous finite staggered magnetization transverse to 
the field. From the theoretical point of view, such an ordered state 
is very well understood
as the result of Bose-Einstein condensation for the $S=1$ triplet 
quasiparticles created by the field \cite{Affleck91,GiamarchiT99, 
Nikunietal00, Matsumotoetal04,Kawashima04,MisguichO04,Wesseletal}, 
with the spontaneous antiferromagnetic ordering corresponding 
to long-range phase coherence of the condensate.

 An alternative way of driving quantum-disordered spin-gap
systems into a spontaneously ordered state is by doping
the magnetic lattice with non-magnetic impurities, as
unambiguosly observed in almost all of the above cited cases,
namely in Haldane chains \cite{Uchiyamaetal99}, coupled spin ladders 
\cite{Azumaetal97}
and 3$d$ weakly coupled dimers \cite{Oosawaetal02}. The effect of 
non-magnetic impurities is the formation of local free $S=1/2$
moments exponentially localized around the impurities. In the 
weakly coupled dimer systems they roughly correspond to unpaired 
spins, while in doped Haldane chains they correspond to
the edge spins of the chain fragments. The overlap between 
two exponentially localized
moments produces an effective coupling between them which
decays exponentially with the impurity-impurity distance 
\cite{SigristF96}.
Despite the fact that the impurities are randomly located, 
such couplings are perfectly unfrustrated and have staggering
signs so that they induce spontaneous long-range 
N\'eel order in the free moments, giving rise to a 
paradigmatic \emph{order-by-disorder} phenomenon \cite{ShenderK91}.

 An intriguing question concerns the fate of the ground state
of the system in presence of \emph{both} site dilution and 
an applied magnetic field. This situation, which is obviously
of direct experimental relevance for all the real systems
cited before, offers in principle the possibility of investigating 
\emph{two} well distinct physical phenomena. On the one side, it is
interesting to study how the order-by-disorder phase
of the site-diluted system is altered and eventually destroyed
by the application of a field, which plays in this
case a disordering role for the system \cite{Mikeskaetal04}. 
At the same time, the spin gap for the clean system can be
orders of magnitude larger than the typical energy scale
of the effective interactions between the $S=1/2$ free moments.
This means that, after destruction of the ordered state of the
free moments, the system can still be driven by an
increasing field through a 
further transition to an ordered state similar to the 
one of the clean case, involving this time the spins which
are far from the impurities. This offers the invaluable
perspective of investigating a phenomenon of Bose-Einstein
condensation in presence of lattice disorder, for which
the appearence of an intermediate
novel disordered phase, the \emph{Bose-glass} phase,
has been predicted long ago \cite{Fisheretal89}, but it has so 
far eluded the experimental observation \cite{Crowelletal95}. 

 In this paper we investigate a specific example of site-diluted
 spin-gap antiferromagnets in a magnetic field, namely 
 a bilayer system in the strong interlayer coupling regime. 
 We choose this specific geometry for two main reasons. 
 One is that a 2$d$ arrangement of weakly coupled dimers
 is the lowest-dimensional structure displaying 
 genuinely ordered phases at $T=0$ 
 and genuine
 order-disorder quantum phase transitions, and 
 at the same time quantum effects remain significant
 due to the reduced dimensionality, in particular
 quantum localization effects which are at the core 
 of the Bose glass phase. One the other hand, a system
 of weakly coupled Heisenberg bilayers is a faithful 
 magnetic model for BaCuSi$_2$O$_6$ \cite{Jaimeetal04}, 
 in which the $S=1/2$ Cu$^{2+}$ ions can be in principle doped with 
 $S=0$ Mg$^{2+}$ or Zn$^{2+}$ ions, thus leading to
 site dilution of the magnetic lattice.
 
\begin{figure}[h]
\begin{center}
\includegraphics[
     width=90mm,angle=0]{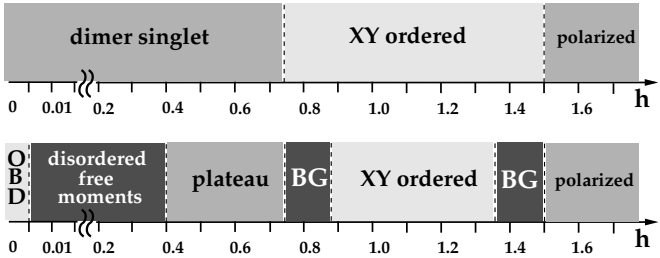} 
\caption{Sequence of ground-state phases in the Heisenberg bilayer in a uniform field $h$
(in dimensionless units, see text). Upper panel: clean case; lower panel:
site-diluted case. The numerical values are referred 
to a bilayer with $J/J'=8$, and, for the lower panel, with 
20\% of vacant sites. For the phases
indicated with an acronym: OBD = order-by-disorder, BG = Bose glass.
The light-shaded regions correspond to \emph{ordered} (and gapless) phases,
the medium-shaded regions to \emph{gapped disorded} phases,
and the dark-shaded regions to \emph{gapless disordered} phases.}
\label{f.phasesketch}
\end{center}
\end{figure}

 Making use of Stochastic Series Expansion (SSE) quantum Monte 
 Carlo (QMC) \cite{SyljuasenS02},
 we can investigate the detailed evolution of the ground 
 state of the doped system upon growing the applied magnetic field.
 The main results of the paper are the following. At a field 
 much smaller than the clean-system gap, the order-by-disorder
 phase is destroyed, but the subsequent field-induced disordered 
 phase, despite having a finite correlation length, is \emph{gapless} 
 as the free moments induced by the impurities are still far from 
 being saturated. We dub this novel quantum phase the 
 \emph{disordered-free-moment} phase. For still higher fields, 
 the full polarization of the free moments leads to a
 \emph{plateau} in the magnetization, and a gap proportional
 to the field opens in the spectrum of the system.
 By further increasing the field towards the lower
 critical value of the clean case, the gap closes again
 and the magnetization starts to rise \emph{exponentially slowly}
 in the field as the system enters a second unconventional 
 quantum phase, namely the aforementioned Bose-glass phase.
 The Bose-glass phase is then turned into a long-range
 ordered (superfluid) phase for the bosonic quasiparticles
 appearing on dimers far from non-magnetic impurities, namely
 triplet quasiparticles for lower field and singlet 
 quasiholes for higher field. An additional Bose-glass phase
 for the quasiholes is then realized before the final 
 high-field phase in which all the spins are fully
 saturated. A sketch of the 
 succession of the phases in the clean \emph{vs.} disordered
 case is shown in Fig. \ref{f.phasesketch}.

 The paper is structured as follows: in Section \ref{sec.intro}
we shortly review the behavior of the Heisenberg bilayer in a field
in absence of disorder; in Section \ref{sec.obd} we 
discuss the order-by-disorder phenomenon in the site-diluted
Heisenberg bilayer in zero field; in Section \ref{sec.results} we 
show our QMC results for the complete field scan for a system 
with 20\% of site
vacancies in the strong interlayer coupling limit; in Section
\ref{sec.DFM} we discuss in details the mechanism of destruction of
the order-by-disorder phase and the emergence of an unconventional 
disordered-free-moment phase; in Section \ref{sec.BG} we discuss the
Bose-glass phase with particular emphasis on the manifestation of 
the rare-event statistics in physical observables as the 
uniform magnetization. Conclusions are drawn in Section 
\ref{sec.conclu}.

 \section{Bilayer Heisenberg antiferromagnet in a field: clean case}\label{sec.intro} 
 
 The Hamiltonian of the $S=1/2$ Heisenberg antiferromagnet on a bilayer 
 reads
 
 \begin{eqnarray}
{\cal H} &=& J' \sum_{\langle ij\rangle} \sum_{\alpha=1, 2}
 {\bm S}_{i,\alpha}\cdot{\bm S}_{j,\alpha}  \nonumber \\
&+& J \sum_{i} 
 {\bm S}_{i,1}\cdot{\bm S}_{i,2}
 - g\mu_B H \sum_{i,\alpha} \epsilon_{i,\alpha} S^z_{i,\alpha}~.
 \label{e.hamilton}
\end{eqnarray}
Here the index $i$ runs over the sites of a square lattice, 
$\langle ij\rangle$ are pairs of nearest neighbors on the 
square lattice, and $\alpha$ is the layer index.
$J$ is the \emph{inter}layer coupling and $J'$ 
the \emph{intra}layer one. Hereafter we will
express the field in reduced units $h = g\mu_B H/J$.

 This model has been extensively studied in the past,
 both analytically \cite{Sommeretal01,Shevchenkoetal00} 
 and numerically \cite{SandvikS94,Shevchenkoetal00}. 
 For $h=0$ the system is in a N\'eel-ordered ground
 state if $g = J/J' < g_c = 2.52..$ ~\cite{SandvikS94,Shevchenkoetal00},
 while for $g > g_c$ the ground state is a total singlet with 
no long-range magnetic order and a finite
gap to all triplet excitations. In the limit of $g \gg g_c$
the ground state can be approximately represented as a collection 
of singlets on the strong interlayer bonds (dimers), so that
the quantum-disordered phase of the model is generally 
indicated as the \emph{dimer-singlet} phase. In what follows
we will focus on a bilayer system with $g = 8$, namely,
in absence of disorder and magnetic field, the system is deep
in the dimer-singlet regime.

Applying a uniform magnetic field (see Fig. \ref{f.phasesketch})
has the effect of lowering the energy of the $S>0$ states aligned
with the field, up to a critical value 
$h^{(0)}_{c1} \approx  1-2/g + O(g^{-2})$~ \cite{Sommeretal01},
corresponding to the singlet-triplet gap $\Delta$,
 where the lowest triplet 
becomes degenerate with the singlet state, and the system develops
a finite uniform magnetization along the field. At this field value 
a dilute gas of hardcore triplet bosonic quasiparticles, corresponding 
to the elementary excitations of the soft triplet mode, appears
in the ground state of the system, and it naturally forms a 
Bose-Einstein condensate  with 
finite superfluid density. The long-range phase coherence 
of the condensate is reflected in the
appearence of a spontaneous \emph{staggered magnetization}
$m_s^{x(y)} =  
 (-1)^{i+\alpha} \langle S_{i,\alpha}^{x} \rangle$ transverse
to the field, with the singular phenomenon of antiferromagnetic
order induced by a uniform field \cite{RoscildeH05}. 
If more bilayers are weakly coupled in a 3$d$ structure,
as in the case of BaCuSi$_2$O$_6$ \cite{Jaimeetal04}, 
the field-induced staggered magnetization persists up to a finite
critical temperature $T_c$, whose scaling with the applied field
is well described by the mean-field theory for a diluted 
repulsive Bose gas \cite{Nikunietal00, Sebastianetal05}.
 
When increasing the field even 
further, the gas of hardcore quasiparticles approaches
the maximum density of one particle per dimer, in which case 
the ordered ground state is better described by a Bose-Einstein 
condensate of singlet quasiholes in the 'triplet sea'. 
Eventually the hole condensate is completely removed by increasing 
the field beyond an upper critical value $h^{(0)}_{c2} =  1+4/g$, at which the 
uniform magnetization reaches its saturation.

\section{Site-diluted Heisenberg bilayer: order by disorder}\label{sec.obd} 

In this section we discuss how the dimer-singlet ground state
of the Heisenberg bilayer is transformed under site dilution 
of the lattice with a concentration $p$ of vacancies well 
below the percolation threshold of the lattice $p^{*}=0.5244(2)$
\cite{RoscildeH05}.
 The Hamiltonian of the site-diluted Heisenberg bilayer reads

 \begin{eqnarray}
{\cal H} &=& J' \sum_{\langle ij\rangle} \sum_{\alpha=1, 2}
\epsilon_{i,\alpha} \epsilon_{j,\alpha} 
 {\bm S}_{i,\alpha}\cdot{\bm S}_{j,\alpha}  \nonumber \\
&+& J \sum_{i} \epsilon_{i,1} \epsilon_{j,2} 
 {\bm S}_{i,1}\cdot{\bm S}_{i,2}
 - hJ \sum_{i,\alpha} \epsilon_{i,\alpha} S^z_{i,\alpha}~.
 \label{e.hamiltondis}
\end{eqnarray}

The variables $\epsilon_{i,\alpha}$ take the values 0 or 1
with probability $p$ and $1-p$ respectively. In this section 
we focus on the case $h=0$.
Starting from $N_{\rm tot}$ spins
the elimination of a fraction $p$ of them leaves
$ p(1-p) N_{\rm tot}$ \emph{unpaired spins} (namely
spins that are missing their dimer partner) and 
$ (1-p)^2 N_{\rm tot}$ spins involved in \emph{intact dimers}.
Normalizing to the number of surviving spins, 
$N = (1-p) N_{\rm tot} $, we obtain a global fraction $p$ 
of unpaired spins. 

 The presence of an unpaired spin on a given site
introduces a signicant local perturbation of the
dimer-singlet state of the system. In fact the 
coupling $J'$ of the unpaired spin to the 
neighboring intact dimers can cause spin flips
of the unpaired spin and simultaneous creation
of a triplet excitation in the intact dimers,
which are then polarized in the original direction
of the unpaired spin. This flip-flop process, albeit  
weak if the energy gap to triplet excitations is large,
effectively \emph{spreads} the overall $S=1/2$ magnetic 
moment of the unpaired spin over the neighboring
intact dimers, within a volume of the order of 
$\xi_0^d$ where $\xi_0$ is the correlation length
in the clean limit. This can be easily seen in first
order perturbation theory, as discussed in the 
Appendix \ref{app.wf}. The large-distance tail of the wavefunction 
of the spread $S=1/2$ free moment in 2$d$, centered around
the $\bm r=0$ site of the unpaired spin, reads 

\begin{equation}
\psi({\bm r}) \approx \frac{J'z}{\Delta}~\frac{e^{-r/\xi_0}}{r} 
\label{e.wf}
\end{equation} 

where $\Delta$ is the triplet gap of the clean system.

Due to their spatial spread, the induced $S=1/2$ magnetic 
moments can overlap and thus effectively interact across 
regions of intact dimers. Given that the wavefunction overlap 
is exponentially decaying with the inter-moment distance,
we should expect the interaction strength $J_{\rm eff}$ to decay 
the same way. The leading contribution to the effective 
interaction can be calculated within second order perturbation
theory \cite{SigristF96,Mikeskaetal04} in a similar 
fashion to the RKKY interaction
between magnetic impurities in a metal;
the details of the calculation in the 2$d$ case are given 
in Appendix \ref{app.wf}. The resulting effective
interaction between two impurities located at sites
$(i,\alpha)$ and $(j,\beta)$ at a distance 
$r = |{\bm r}_i-{\bm r}_j|$ is 
${\cal H}_{\rm eff} = J_{\rm eff} {\bm S}_{i,\alpha}\cdot {\bm S}_{j,\beta}$,
where, in the large-$r$ limit
\begin{equation}
J_{\rm eff} (i,j;\alpha,\beta) \approx (-1)^{\sigma}~
\frac{J_1}{r} ~e^{-r/\xi_0}
\label{e.jeff}
\end{equation}
with 
\begin{equation}
J_1 =  \frac{(zJ')^2}{4\pi\Delta}.
\end{equation}
and $\sigma = x_i + y_i + x_j + y_j +\alpha+\beta$.
 Due to this staggering factor, such interactions form 
 an \emph{unfrustrated} network 
 which induces long-range antiferromagnetic ordering
 of the free moments. The resulting ordered phase will be
 denoted in the following as the \emph{order-by-disorder
 phase}.  

 The order of magnitude of the interaction is  
 mainly set by the ratio $(J')^2/\Delta$. Given that 
 $\Delta \sim J$, this means that in the strong
 interlayer coupling regime $g\gg 1$ the order-by-disorder
 phenomenon has a characteristic energy scale for
 excitations which can be orders of magnitude smaller than 
 the energy ($\sim J$) for the excitations living far away from the 
 impurities, namely in locally clean regions of the system. 
 A more precise estimate of the typical energy scale 
 for the effective interactions between
 the free moments is obtained by averaging Eq. \ref{e.jeff}
 over its probability distribution.
 Here we simply consider pairs of closest-neighboring
 spins, discarding longer-range pairs due to the 
 exponential decay of their mutual coupling.  
 In the continuum limit,
 the probability for an unpaired spin to have its
 closest neighbor within a disk of radius $R$ is given by
 \begin{equation}
 P(R) = 2p~e^{-2p\pi R^2}
 \end{equation}
 This probability distribution is normalized on the 
 infinite disk. The corresponding probability distribution 
 for the $J_{\rm eff}$ couplings (here taken without
 the staggering sign) is then
 \begin{eqnarray}
 \tilde{P}(J_{\rm eff}) &=& \frac{P(R(J_{\rm eff}))}
 {|J_{\rm eff}'(R(J_{\rm eff}))|} \nonumber \\
 &=& \frac{4\pi p R \xi_0}
 {J_{\rm eff} (1+\xi_0/R)}~ e^{-2p\pi R^2} \Theta(J_1 e^{-1/\xi_0}-J_{\rm eff})~~~~~~~~
 \end{eqnarray}
 where the upper cutoff on $J_{\rm eff}$ descends from
 the lower cutoff on the inter-moment distance $R = 1$
 (corresponding to one lattice spacing), and it is necessary 
 to regularize the distribution. The strength $J'$ of the 
 inter-moment coupling for $R=1$ is actually underestimated 
 by the asymptotic formula Eq.~(\ref{e.jeff}).
 To take this into account one can in principle 
 introduce the corrected distribution
 \cite{Mikeskaetal04} $\tilde{\tilde{P}}(J_{\rm eff})$:
 \begin{equation}
 \tilde{\tilde{P}}(J_{\rm eff}) = \tilde{P}(J_{\rm eff}) 
 + 4p ~\delta(J_{\rm eff} - J').
 \end{equation}
 We notice however that unpaired spins lying at the 
 distance of a single lattice spacing will have a
 strong tendency to form a singlet state and therefore
 not to participate to the long-range N\'eel ordered
 state of the system \cite{Mikeskaetal04}
 (see also the discussion in 
 Sec. \ref{sec.DFM}). Therefore, when estimating
 the characteristic energy scale associated with
 the N\'eel ordered state of the free moments, we
 can discard the singular part of the distribution,
 obtaining therefore the following average 
 effective coupling $\langle J_{\rm eff} \rangle$:
 \begin{equation}
 \langle J_{\rm eff} \rangle =
 \int dJ_{\rm eff} ~J_{\rm eff} \tilde{P}(J_{\rm eff}) 
 \approx 2\sqrt{2} \pi ~p ~J_1 e^{-1/\xi_0}.
 \label{e.jeffav}
 \end{equation}

\section{QMC method and results}\label{sec.results}

\begin{figure}[h]
\begin{center}
\includegraphics[
     width=90mm,angle=0]{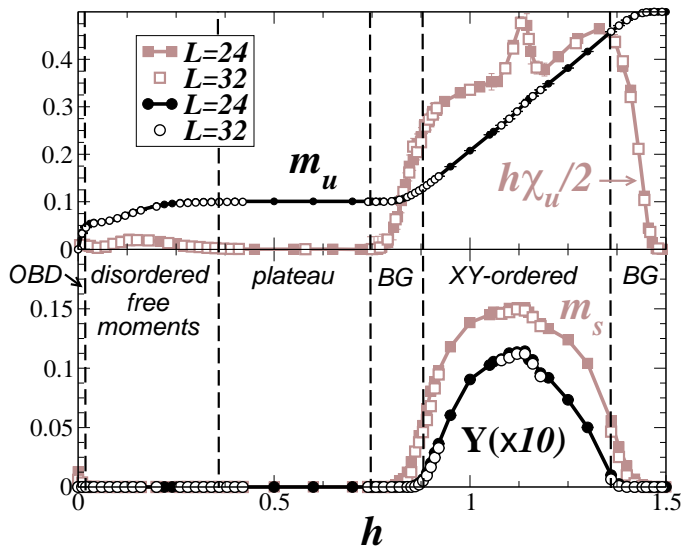}
\caption{Zero-temperature field scan for the 
site-diluted Heisenberg bilayer with $g=8$ and $p=0.2$.}
\label{f.hscan}
\end{center}
\end{figure}  

 In this section we present our numerical results for the 
 Hamiltonian Eq. \ref{e.hamiltondis} in a finite field $h$.
 The method we use is the Stochastic Series Expansion (SSE)
 QMC based on the directed-loop algorithm 
 \cite{SyljuasenS02}. Despite using
 a finite-temperature QMC approach, we can systematically
 study the ground-state physics of the model by 
 efficiently cooling the system to its physical $T=0$
 limit via a successive doubling of the inverse temperature
 \cite{Sandvik02}. $L\times L \times 2$ lattices up 
 to $L=40$ have been considered. Here the size refers to the 
 lattice \emph{before} depletion. All the results for $p>0$
 have been averaged over at least 200 disorder realizations. 
 
  Fig. \ref{f.hscan} shows the $T=0$ field dependence
  of the most relevant observables for a site-diluted
  Heisenberg bilayer in the strong interlayer coupling 
  regime $g=8$,
  and with depletion $p=0.2$. Plotted are the uniform 
  magnetization along the field $m_u = \langle S_i^{z}\rangle$,
  the uniform susceptibility $\chi_u = dm_u/dh$, the staggered
  magnetization transverse to the field $m_s = 1/4 \sum_{\alpha\beta} 
\sqrt{ (-1)^{L/2+\alpha+\beta} \langle S_{i,\alpha}^{x} 
S^{x}_{i+L/2,\beta} \rangle}$ and the spin stiffness (superfluid
density) $\Upsilon = k_B T/(2J) ~\langle W_1^2 + W_2^2 \rangle$,
where $W_{1,2}$ are the winding numbers in the two lattice 
directions of the worldlines
appearing in the SSE representation of the quantum partition
function \cite{PollockC87,Sandvik97}. 
 
 It is straightforward to observe that non-magnetic impurities
introduce an extremely rich field dependence of the magnetic 
observables, which differs substantially from the one observed
in the clean case \cite{Sommeretal01,RoscildeH05}. Here we describe the 
alternation of phases induced by the field, postponing the 
details of the two novel quantum-disordered phases (disordered-free-moments
and Bose-glass phase) to the following sections.

In zero field, long-range 
order is expected due to an order-by-disorder mechanism, as 
discussed in Sec. \ref{sec.obd}. To estimate the average energy scale 
associated with the ordered phases through Eq. \ref{e.jeffav}
we need the information the correlation length $\xi_0$ and the 
gap $\Delta$ in the clean case $p=0$. 
From QMC simulations for the
$p=0$ case we obtain $\xi_0 \approx 0.5$, through
second moment estimation \cite{Cooperetal82}. Moreover we
can estimate the gap through the critical field $h_{c1}^{(0)}$
that induces long-range antiferromagnetic order in the 
clean system. We obtain $h_{c1}^{(0)} = \Delta/J = 0.745(2)$
through the scaling of the uniform magnetization $m_u$,
which becomes finite at the critical point. 
The resulting estimate for  $\langle J_{\rm eff} \rangle$ is then: 
$\langle J_{\rm eff} \rangle \approx 6\times 10^{-3} J$. 

\begin{figure}[h]
\begin{center}
\includegraphics[
     width=80mm,angle=0]{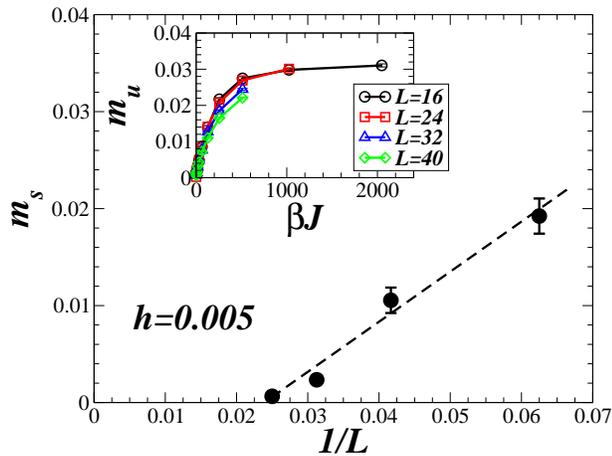}
\caption{Scaling of the staggered magnetization in the site-diluted
Heisenberg blayer with $p=0.2$, $g=8$ and at $h=5\times 10^{-3}$. 
The dashed line is a guide to the eye. 
\emph{Inset}: uniform magnetization as a function of the inverse
temperature for the same model parameters. Notice that, 
despite $m_u$ does not show full temperature saturation
for the bigger lattice sizes, 
it is evident that it scales towards smaller values
for increasing $L$, so that in the thermodynamic limit
it will converge to a value well below the plateau one 
$m_u < pS=0.1$.}
\label{f.h0.005}
\end{center}
\end{figure}  

In presence of a field, the zero-field N\'eel-ordered state
of the free moments 
will turn into a \emph{canted XY-ordered} state, and eventually
the field will destroy the antiferromagnetic order of the
free moments leading to their full polarization. Nonetheless,
due to the peculiar features of the long-range interactions
between the free moments, the canted XY-ordered phase and 
the fully polarized phase are \emph{not contiguous}, but
they are separated by an intermediate, novel phase.

The extremely small value for $\langle J_{\rm eff} \rangle$
compared with the gap in the clean case
immediately suggests that an equally small field can 
suppress long-range antiferromagnetism. Estimating this
field numerically is a formidable task, given that, to observe
the $T=0$ physics on reasonably large sizes, we need to 
be at $k_B T \ll \langle J_{\rm eff} \rangle$. Our QMC simulations
show unambiguously that no long-range order survives the extrapolation 
to the thermodynamic limit down to a field $h=0.005$, as shown in Fig.
\ref{f.h0.005}.  Therefore we set the value $h=0.005$ as an upper bound 
for the destruction of the order-by-disorder phase.

Remarkably, the small field that destroys the order-by-disorder phase 
is able to only \emph{partially} polarize the free moments localized
around the unpaired spins, as shown in the inset of Fig. \ref{f.h0.005}. 
In fact, immediately after the destruction
of long-range order, the magnetization remains well below the 
saturation value for the free moments, corresponding to 
$m_u^{*} = pS = 0.1$. The state of partial polarization of the
free moments persists up to a field $h=h_{\rm plateau} \approx 0.36$
at which the magnetization reaches a plateau corresponding
exactly to the value $m_u^{*}$. This means that, for 
$0.005 \lesssim h < h_{\rm plateau}$, the free moments still
preserve a finite projection on the $xy$ plane transverse to
the field, and that the transverse spin components are 
\emph{quantum disordered}. This quantum disordered state
is \emph{gapless}, given that the magnetization continues
to grow with the field. 
To our knowledge this phase has no analog in what has been
observed so far in clean systems, and 
we therefore dub it as \emph{disordered-free-moment phase}. 
A possible scenario for 
the mechanism leading to quantum disorder in this phase is 
provided in Sec. \ref{sec.DFM}.

 For $h>h_{\rm plateau}$ the system has fully polarized 
 free moments, and it acquires a gap to all triplet
 excitations, corresponding to a vanishing of the uniform 
 susceptibility. This state persists over a quite sizable
 field range: in this field interval the dynamics
 of the free moments is completely quenched by the 
 fields, but intact dimers lying away from unpaired spins
 have a local gap which is still larger than the field.
 Therefore, by dynamically eliminating the extra degrees of freedom 
 introduced by the unpaired spins, the field essentially
 restores a gapped disordered state which is the "dirty"
 counterpart to the dimer-singlet state in the clean
 limit. Interestingly, while the gap for the excitations
 of the free moments \emph{increases} linearly with the field,
 the one for the excitation of the intact dimers 
 in clean regions \emph{decreases}, and eventually closes
 in a fashion similar to that of the clean limit, with
 the appearence of a dilute gas of triplet quasiparticles.
 The field value at which the gap to clean-region excitations 
 closes is necessarily the same as the
 lower-critical field in the clean limit $h_{c1}^{(0)}$.
 In fact, in the thermodynamic limit there exists always
 an arbitrarily big clean region with 
 probability 
 \begin{equation}
 P(l) \sim \exp(-2|\ln(1-p)|l^{2})
 \label{e.Pl}
 \end{equation}
 where $l$ is 
 its characteristic linear size. This region,
 having an arbitrarily big size albeit with infinitesimal probability,
 is arbitrarily close to a clean system, and for $h \geq h_{c1}^{(0)}$
 its local gap must close, thus accepting the appearence of 
 the first triplet quasiparticles. This is fully consistent
 with our QMC results, where we observe a revival of the
 magnetization process after the plateau phase at
 $h\approx 0.76$  for lattice sizes up to $L=32$, which 
 is already in good agreement with $h_{c1}^{(0)}\approx 0.745$.
 It is important to point out that any finite-size estimate
 of the closure of the gap in a disordered system is
 an \emph{upper bound} to the actual value, due to the 
 fact that a rare clean region can be at most as big
 as the size of the entire finite-size system. 
 
  A fundamental difference with the closure of
  the gap in the clean case is provided by the fact that the
  first triplet quasiparticles appearing in the largest clean regions 
  of the system are \emph{localized} in such regions, and are not
  able to coherently propagate throughout the system. In fact, fully  
  polarized free moments with a finite gap to spin-flips 
  act as almost impenetrable scatterers for the triplet quasiparticles.
  On the other hand, intact dimers close to impurities have a lower coordination
  to the rest of the system and therefore they feature a local gap which is 
  bigger than the one of the clean two-dimensional system, $h_{c1}^{(0)}J$,
  and closer to that of a single dimer, $J > h_{c1}^{(0)}J$. 
  This means that they also act as energy barriers for 
  triplet quasiparticles created in the clean
  regions. If $h\gtrsim h_{c1}^{(0)}J$, the gas of triplet quasiparticles
  is extremely dilute, and we can neglect interactions among them. 
  The state of the triplet quasiparticles at low filling in the site-diluted
  weakly coupled dimer system is therefore analogous to the 
  ground state of an Anderson problem in two dimensions, which is always
  \emph{quantum localized} regardless of the strength of disorder
  and of the dimensionality. Therefore, after the closure of the 
  gap in the clean regions, the system of triplet quasi-particles
  form an Anderson-insulating state, which is usually
  denoted as Bose glass for interacting bosons \cite{Fisheretal89}.
  Such state has \emph{no condensate} and \emph{no superfluid fraction},
  corresponding to a vanishing transverse magnetization $m_s$
  and to a vanishing spin stiffness $\Upsilon$, as clearly shown
  by our QMC results. Yet, it is a compressible state, namely
  it has a vanishing particle gap, corresponding to a finite
  susceptibility $\chi_u$ in the magnetic language.
  
  \begin{figure}[h]
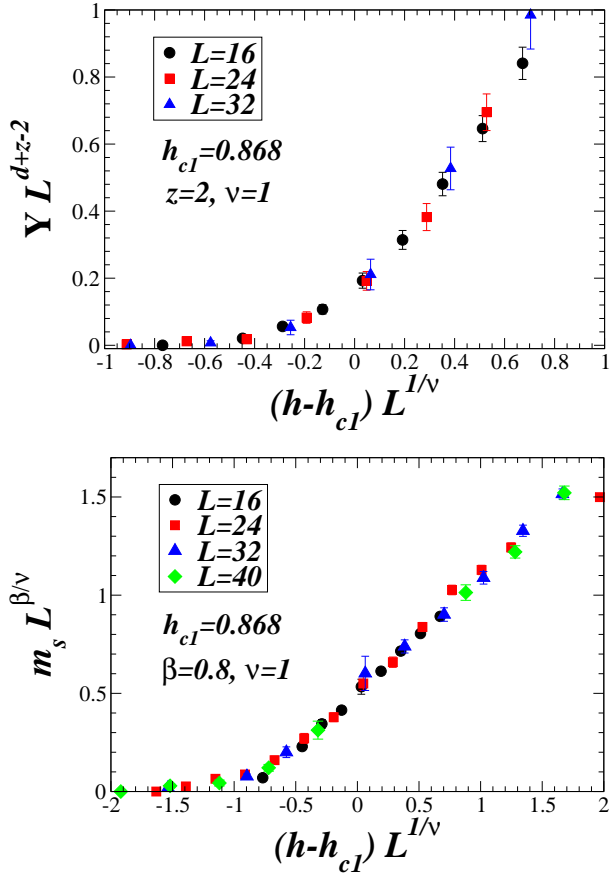

\begin{center}
\includegraphics[
     width=80mm,angle=0]{fig4a.eps}
     \vskip .3cm
 \includegraphics[
     width=80mm,angle=0]{fig4b.eps}         
\caption{Scaling of the superfluid density $\Upsilon$
and of the order parameter $m_s$ at
the Bose-glass-to-superfluid transition for the site-diluted
bilayer Heisenberg antiferromagnet with $g=8$ and $p=0.2$.}
\label{f.YYscaling}
\end{center}
\end{figure}  
    
   As the field is increased above $h_{c1}^{(0)}$, the density of
   bosons increases, and the extent of locally gapless regions 
   increases as well. The hard-core repulsion between the quasiparticles
   and their increased chemical potential eventually leads to a 
   localization-delocalization transition, corresponding to the 
   onset of superfluidity ($\Upsilon > 0$) and to the appearence 
   of long-range phase order ($m_s>0$). The Bose-glass-to-superfluid
   quantum phase transition happens at the critical field $h_{c1}= 0.87(2)$,
   which we estimate through the study of the scaling of the correlation length,
   superfluid density and staggered magnetization. The scaling theory
   of Ref. \onlinecite{Fisheretal89} formulates specific predictions
   for the dynamical critical exponent at the transition, namely 
   $z=d=2$. This exponent appears in the quantum-critical scaling form
   of the superfluid density
   \begin{equation}
   \Upsilon = L^{-(d+z-2)} F_{\Upsilon}[L^{1/\nu}(h-h_{c1})].
   \end{equation}
   where $\nu$ is the critical exponent of the correlation length.
   Moreover we also consider the quantum critical behavior of
   the order parameter $m_s$, in terms of the scaling form:
   \begin{equation}
   m_s = L^{-\beta/\nu} F_{m_s}[L^{1/\nu}(h-h_{c1})].
   \end{equation}
   Fig. \ref{f.YYscaling} shows the plots of the rescaled superfluid
   density $\Upsilon L^{d+z-2}$ and of the 
   rescaled order parameter $m_s L^{\beta/\nu}$ as a function of 
   the rescaled distance from the critical point $L^{1/\nu}(h-h_{c1})$.
   The predicted $z=2$ provides a very good collapse
   of the different $\Upsilon$ curves at different sizes
   together with the exponent $\nu\approx 1.0(1)$, satisfying
   the Harris criterion \cite{Chayesetal86} $\nu \geq 2/d$ , and
   with $h_{c1}\approx 0.868$, consistent with the
   otherwise estimated $h_{c1} = 0.87(2)$. Moreover 
   the scaling study of the staggered magnetization
   provides an estimate for the exponent $\beta=0.8(1)$. 
   The result for the dynamical critical exponent 
   remarkably shows that the site-diluted Heisenberg bilayer 
   fully realizes the theoretical picture of the transition 
   from a Bose-glass state to a superfluid. In $d=2$ the dynamical 
   critical exponent is accidentally unchanged
   with respect to its clean value $z=2$ ~\cite{Sachdev99},
   but the other estimated exponents strongly differ from their mean-field
   values $\beta=1/2$ and $\nu=1/2$ which should hold
   in the clean limit, given that $d+z=4$ is the upper
   critical dimension. 
   
    In the superfluid phase we notice that all the quantities
    shown in Fig. \ref{f.hscan} display a feature at a field
    $h$ corresponding approximately to half filling of the
    intact dimers with triplet quasiparticles, namely to a
    magnetization $m_u = S(1+p)/2=0.3$. The uniform magnetization
    shows a kink, which reflects in a peak of the
    uniform susceptibility; the staggered magnetization $m_s$
    and the superfluid density $\Upsilon$ also show a kink.
    This singular behavior is not observed at all  in the clean system 
    \cite{RoscildeH05} at half filling. 
    Such a feature is probably to
   be attributed to weakly coordinated dimers or small clusters
   of dimers present in the system, whose response to a field is 
   step-like when the field exceeds their local gap. See also
   Sec. \ref{sec.BG} for further discussion.
   
   Given the hardcore nature of the triplet quasiparticles, a filling of 
   at most a triplet quasiparticle per intact dimer can be reached
   by the system. When the density of quasiparticles gets close
   to its maximum value, the system can be more conveniently regarded
   as a dilute gas of \emph{singlet quasiholes} in the triplet sea,
   living on intact dimers which are not fully polarized. Interestingly,
   the intact dimers belonging to clean regions can be fully polarized
   only by a field close to the saturation field of the clean limit 
   $h \lesssim h_{c2}^{(0)}= 1+4/g$. On the contrary, intact dimers which are
   in regions of lower local coordination are more easily polarized,
   because their local saturation field is closer to that of an
   isolated dimer, $h=1$. This means that singlet quasiholes
   get gradually expelled from regions close to the impurities
   and get \emph{localized} in the clean regions, analogously
   to what had happened to the triplet quasiparticles for 
   $h \gtrsim h_{c1}^{(0)}$. The superfluid state of the singlet
   quasiholes gets therefore destroyed by the field, and the 
   system undergoes a \emph{second} superfluid-to-Bose-glass transition
   at a critical field $h_{c2} = 1.36(2) < h_{c2}^{(0)} = 1.5$.
   After a second extended Bose-glass phase for $h_{c2} < h < h_{c2}^{(0)}$,
   the system of triplet quasiparticles reaches unitary filling,      
    and at this point the system becomes a \emph{band insulator}
   (or, alternatively, a Mott insulator with infinite on-site repulsion).
    Such state corresponds to the saturation of the 
    magnetization, $m_u = S = 1/2$. It is important to point out that,
    in the thermodynamic limit, full saturation is only attained at
    the \emph{clean} critical field $h_{c2}^{(0)}$ and not before:
    in fact, in an infinite system there exists always an arbitrarily 
    big clean region whose local saturation field is arbitrarily close 
    to that of the perfectly clean system.

\section{Field destruction of order by disorder: the disordered-free-moment
phase}\label{sec.DFM}

\begin{figure}[h!]
\begin{center}
\includegraphics[
     width=90mm,angle=0]{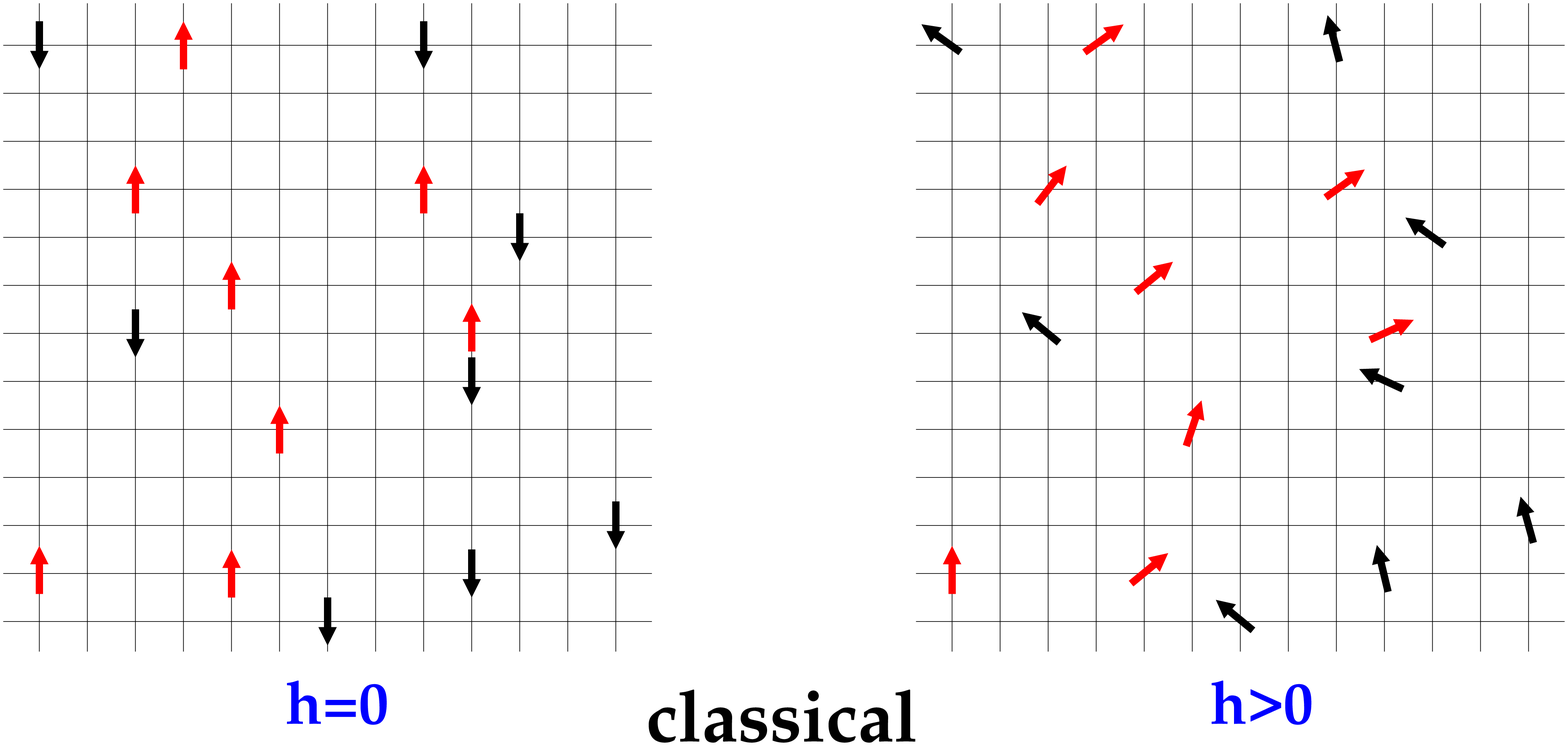}
\vskip 1cm
\includegraphics[
     width=90mm,angle=0]{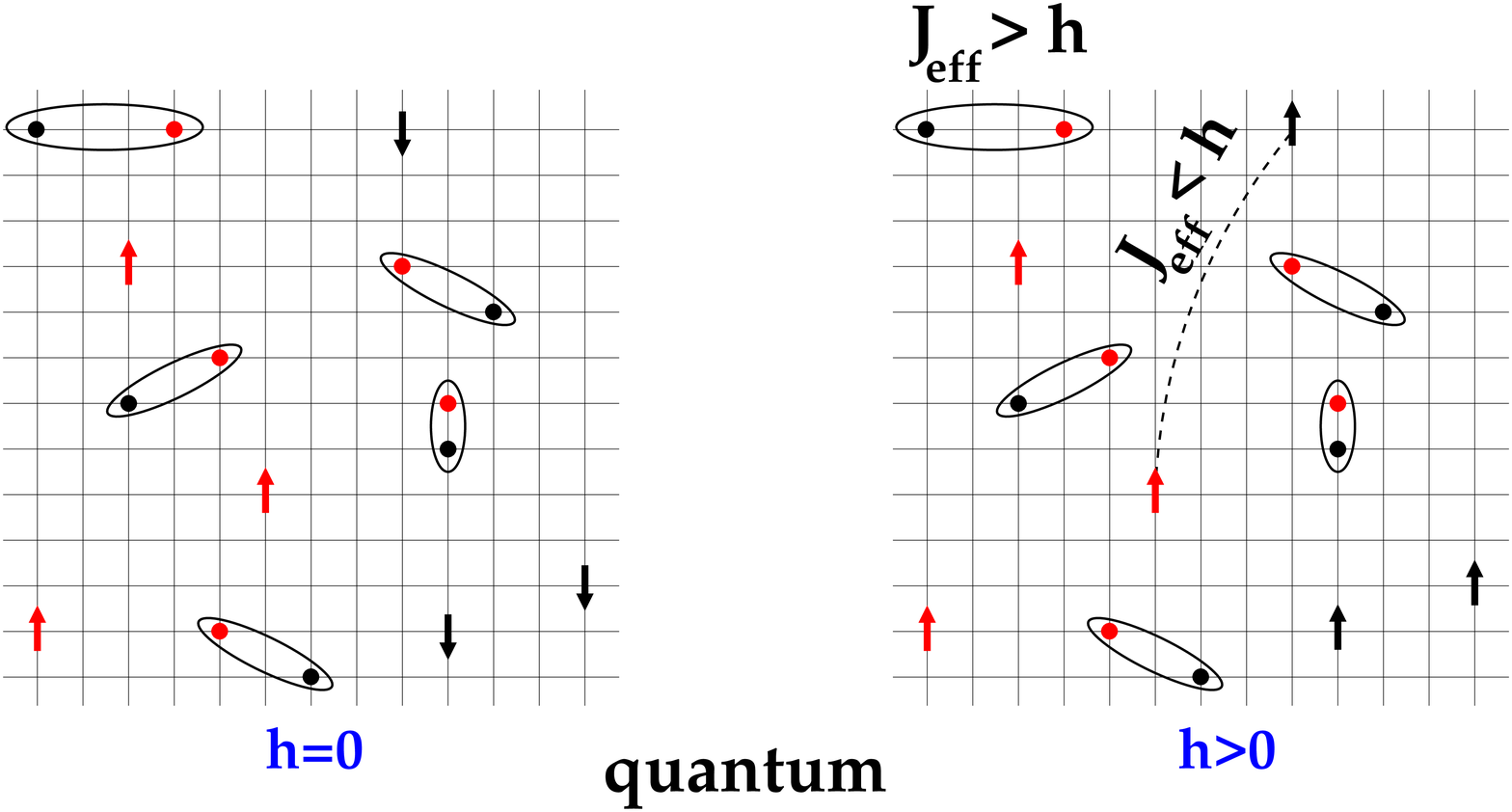}     
\caption{Canted antiferromagnetic ordering of the free moments 
in the classical limit (\emph{upper panel}) \emph{vs.} quantum 
disordered-free-moment phase (\emph{lower panel}). The different
color coding of the spins denotes the two different sublattices.
In the lower panel, ellipses surrounding two sites denote singlet
states, and the dashed line denotes
an unsatisfied antiferromagnetic bond which is overcome by the field.
The state sketched in the lower-right panel has no long-range 
antiferromagnetic order, and it corresponds to the 
disordered-free-moment phase.}
\label{f.DFM}
\end{center}
\end{figure}  
 
  In this section we focus on the field destruction of the long-range
  ordered phase of the free moments. As pointed out in the previous section, 
  the resulting disordered-free-moment phase has the markedly unconventional 
  feature of displaying a gapless spectrum and absence of spontaneous 
  long-range order. Here we propose a physical scenario for the explanation 
  of such phase. 
  
   To simplify the picture of the free moments, we imagine them
   to be fully equivalent to a system of randomly located spins 
   on a square lattice and interacting
   with exponentially decaying Heisenberg couplings. Fig. \ref{f.DFM} sketches
   such system in the simpler case of a single layer - the case of a 
   bilayer is anyway completely analogous. In the classical
   limit $S\to \infty$, the system of randomly located spins in 
   a weak enough magnetic field has a canted antiferromagnetic 
   ground state (Fig. \ref{f.DFM}, upper panel), in which the 
   spin components transverse to the field are staggered
   according to a 2-sublattice structure. Locally it might 
   happen that isolated spins minimize their energy by 
   fully aligning with the field and losing therefore 
   their transverse components. Nonetheless, the remaining 
   clusters of partially
   polarized spins, even if separated by fully polarized ones,
   will preserve the long-range antiferromagnetic order 
   of their transverse components because they are 
   directly coupled through long-range interactions that
   go across the polarized regions. This means
   that, in the classical limit, the system has long-range 
   order up to the field that roughly equals the strongest coupling 
   $(J_{\rm eff})_{\rm max}$ and thus 
   polarizes all the spins.
   
   Quantum mechanically it is easy to imagine a substantially 
   different ground state. A fundamental phenomenon introduced
   by quantum fluctuations is the formation of local singlets
   between spins that lie close to each other on different sublattices 
   and are therefore strongly coupled through $\tilde{J}_{\rm eff}\sim J'$. 
   If all the other spins are
   sufficiently far apart, it is evident that the close-lying spins
   will have a strong tendency to form a singlet as long as their
   singlet-to-triplet gap is much larger than the sum of the interactions
   with all the other spins, $\tilde{J}_{\rm eff} \gg \sum' J_{\rm eff}$
   where $\sum'$ runs over all spins but the one lying close. The formation
   of a local singlet decouples the spins involved in it 
   from the rest of the system.
   If two other spins exist nearby which are lying on different sublattices,
   are close enough to each other (but not as close as the previous two ones) 
   and far from the others, they will also have a tendency to form a singlet
   with a smaller gap, etc. In one-dimensional bond disordered 
   antiferromagnets this argument leads to the prediction of  
   a gapless \emph{random-singlet phase} \cite{Fisher94} without 
   long-range order, where singlets exist to all energy scale. 
   Interestingly, in two dimensions the above reasoning is only
   valid for close-lying spin pairs, namely only over a short length 
   scale. Due to the higher coordination in a two-dimensional lattice,
   when two spins are sufficiently far from each other the long-range 
   singlet formation becomes unfavourable with respect to the
   appearence of long-range N\'eel order. Indeed, in 
   Ref. \onlinecite{Laflorencieetal04} it has been shown that, in zero-field, a model 
   of randomly located spins with exponentially decaying couplings always 
   displays long-range order regardless of the concentration of the spins.  
   The ground state of the system is 
   evidently very inhomogeneous, with close-lying spins forming approximately
   singlets and partially isolated spins being instead involved in a 
   long-range ordered N\'eel state. A notable feature of this ground
   state is that short-range singlets have a triplet gap
   which can be substantially larger than the typical energy
   associated with the N\'eel state formed by the other spins. 
   Therefore not only the state is highly inhomogeneous, but 
   it also displays a broad range of energy scales. 
   
    Accepting this sketchy view of the ordered state of 
   randomly located spins in zero field, it is trivial to argue what happens
   in presence of a field. Due to the large spread in energy
   scales, the response to an applied field will be very
   inhomogeneous. As in the classical case, the partially
   isolated spins will be the first to be polarized by the
   field, thus minimizing their Zeeman energy. But, according
   to what noticed before, these spins are also the ones 
   that are involved in the long-range ordered state. If there
   is a net energy separation between the characteristic energy
   of the N\'eel state $J_{\text{N\'eel}}$ for these
   spins and the characteristic energy of the singlet states of the 
   close-lying spins $\tilde{J}_{\rm eff}$, there exists also
   a field value $J_{\text{N\'eel}}/J < h <  \tilde{J}_{\rm eff}/J$
   which destroys the N\'eel state of the isolated spins
   by polarizing them, but \emph{does not} polarize the close-lying 
   spins. This field would then correspond to the one which drives
   the order-by-disorder phase into the disordered-free-moment
   phase. The surviving unpolarized singlets are responsible
   for the magnetization not being fully saturated. Their
   maximum spatial range is fixed by the field value, since 
   singlets cannot survive when the coupling
   energy $J_{\rm eff}$ is overcome by the Zeeman energy $hJ$.
   Therefore we expect the antiferromagnetic correlations
   in this phase to be \emph{short-ranged}, as our QMC data 
   seem to suggest, although a direct analysis of this point
   is deferred to future work \cite{Rongetal06}. Such a quantum-disordered
   state is clearly gapless, given that the surviving singlets 
   have in-field gaps that roughly span the continuous interval
   $[0,J'(1-h)]$. 
 
    A recent mean-field study of the field destruction in 
    site-diluted weakly-coupled dimer systems is presented
    in Ref. \onlinecite{Mikeskaetal04}. This study does
    not report any disordered-free-moment phase between 
    the order-by-disorder phase and the plateau one, although
    the mechanism of local singlet formation is clearly mentioned,
    and the spins involved in local singlets are explicitly
    excluded from the mean-field treatment and from the 
    disorder average (compare the discussion in Sec. \ref{sec.obd}). 
    Therefore the
    conclusions of Ref. \onlinecite{Mikeskaetal04} only
    apply to the free moments participating in the N\'eel state
    in zero field and do not exclude the existence of 
    a disordered-free-moment phase as the one we have 
    described.

\section{Bose glass phase and rare-event statistics}\label{sec.BG}
 
 \begin{figure}[h!]
\begin{center}
\includegraphics[
     width=75mm,angle=0]{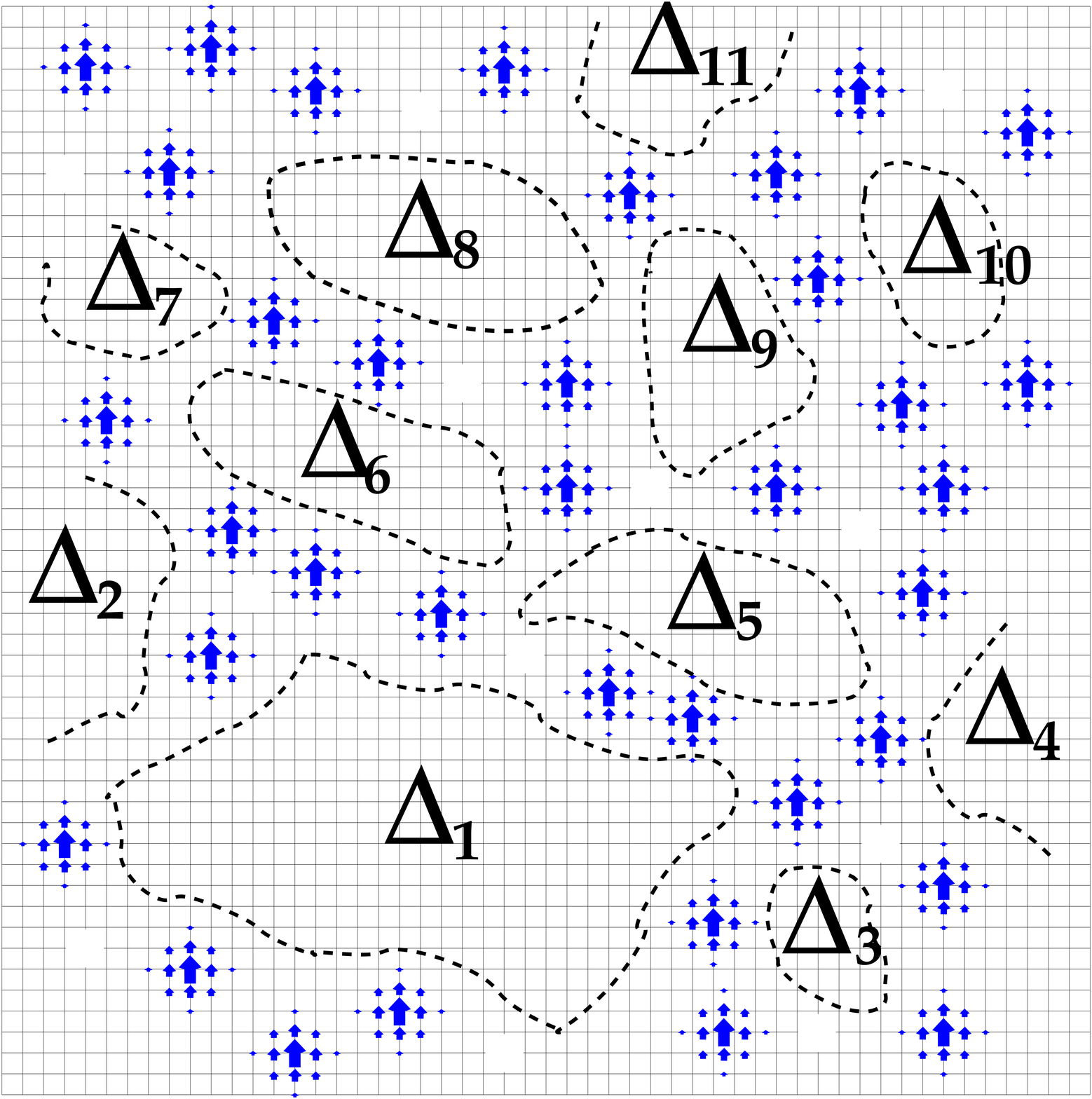}
\vskip 1cm
\caption{Sketch of the \emph{local-gap model} for the
relevant degrees of freedom close to the clean 
critical fields $h_{c1}^{(0)}$ and $h_{c2}^{(0)}$
in the site-diluted Heisenberg bilayer with strong
interlayer coupling. Each 
site of the square lattice corresponds to a dimer 
perpendicular to the plane of the figure; the
small arrows represent the free $S=1/2$ moments 
exponentially localized around an unpaired spin,
and fully polarized by the applied field, while
vacancies correspond to missing dimers. The 
clean regions in the system are highlighted,
and we associate to each of them a local gap 
corresponding to the gap of a finite cluster
with the same size.}   
\label{f.localgap}
\end{center}
\end{figure}

  In this section we discuss in details the nature of the Bose-glass
  phase and of its unconventional magnetization behavior. 
  As already pointed out in Sec. \ref{sec.results}, the characteristic
  feature of this phase is the appearance of quantum-localized
  triplet quasiparticles (singlet quasiholes) in rare clean regions, 
  corresponding to 
  locally magnetized (not fully polarized) intact dimers. 
  Here we assume that
  clean regions hosting
  quantum-localized quasiparticles are completely uncorrelated
  from each other, which is consistent with the picture that
  the quasiparticles are unable to propagate coherently throughout
  the system.   
  From the point of view of the
  response to a magnetic field, this means that each clean region behaves
  independently from the others, and it essentially 
  behaves as a finite-size cluster 
  with characteristic linear size $l$ following the exponentially
  decaying distribution Eq.~(\ref{e.Pl})
  characteristic of the geometrical statistics of
  rare regions. In particular, the response to a magnetic
  field is dictated by the spectral properties of the 
  cluster, namely by the value of the local gap $\Delta(l)$.
  Treating the response of the system as the sum of 
  independent responses of different finite-size
  clusters, governed by their local gap, represents the 
  core assumption of a \emph{local-gap model}
  for the Bose-glass phase (see Fig. \ref{f.localgap}).
  
  First we consider the low-field Bose-glass phase, namely
  the one occurring for $h\gtrsim h_{c1}^{(0)}$. In this
  field regime, a cluster with a gap  $\Delta(l)$ will only respond
  to a field larger or equal to the gap, namely it 
  develops a total uniform magnetization of the type
  \begin{equation}
  M_l(h) = [h-\Delta(l)]^{\gamma} ~\Theta[h-\Delta(l)].
  \end{equation}
  where we approximate the stepwise magnetization curve of 
  a finite cluster through a power law
  with an exponent $\gamma$ whose knowledge
  is not essential for our conclusions, although
  we expect $\gamma=1$ for clusters with a bilayer
  structure \cite{Sommeretal01}.
  For $h_{c1}^{(0)} < h < h_{c2}^{(0)}$ the infinite clean
  system is gapless, so in this field range the gap on finite
  clusters is purely a \emph{finite-size} one.
   We can  
  assume that the finite-size correction to the \emph{zero-field}
  gap of the infinite system, $\Delta(\infty) = h_{c1}^{(0)}$,
  scales with inverse of the cluster volume,
  as expected in Heisenberg antiferromagnets undergoing 
  spontaneous symmetry breaking \cite{Andersonbook}
  \begin{equation}
  \Delta(l) \approx h_{c1}^{(0)} + \frac{c}{l^2} . 
  \end{equation}
  The above relation allows us to relate the probability distribution
  of local gaps to that of the cluster sizes, Eq.~(\ref{e.Pl}):
  \begin{eqnarray}
  \tilde{P}(\Delta) &=& \frac{P[l(\Delta)]}{|\Delta '[l(\Delta)]|} \nonumber \\
  &\sim& 
  \left(\Delta -h_{c1}^{(0)}\right)^{3/2} \exp\left[ -\frac{2|\ln(1-p)|}
  {c \left(\Delta -h_{c1}^{(0)}\right)}\right].~~~~~~
  \end{eqnarray}
  As expected, this distribution attributes an exponentially vanishing
  probability to nearly gapless clusters, reflecting the exponentially
  rare occurrence of large clean regions. 
  
  This local-gap model allows us then to extract the total magnetization
  of the clean clusters as the sum of the magnetizations of the 
  individual clusters:  
  \begin{eqnarray}
  m_u(h) - pS&=& \frac{1}{N(1-p)} \sum_{\rm clusters} M_{\rm cluster}(h) \nonumber \\
  &=& \int d\Delta~(h-\Delta)^{\gamma} ~\Theta(h-\Delta) \tilde{P}(\Delta).~~~~~~ 
  \end{eqnarray}
  Here the magnetization of the clean clusters is expressed as the
  global magnetization $m_u(h)$ minus the saturated magnetization of the 
  free moments, $pS$, corresponding to the magnetization plateau.
  
  The field dependence of the magnetization can be determined to leading
  order through a saddle-point approximation of the above integral over
  the probability distribution $\tilde{P}(\Delta)$, which, for
  $0< h-h_{c1}^{(0)} \ll 1$, yields the following prediction:
  \begin{equation} 
   m_u(h) - pS \sim \exp\left[ -\frac{2|\ln(1-p)|} 
  {c\left(h -h_{c1}^{(0)}\right)}\right] ,
  \label{e.expmagn}
   \end{equation}
  namely an \emph{exponentially slow} magnetization, which reflects
  directly the statistics of the rare clean regions. 
  
   A completely analogous treatment of the field-dependence of the 
   magnetization can be used for the high-field Bose-glass phase,
   $h \lesssim h_{c2}^{(0)}$. In this case we introduce the conjugate 
   magnetization $\tilde{m}_u = S - m_u$ and the conjugate field
   $\tilde{h} = h_{c2}^{(0)} - h$, such that $\tilde{m}_u$ grows
   monotonically with $\tilde{h}$. The \emph{in-field} finite-size gap of the 
   clusters for $h \lesssim h_{c2}^{(0)}$ is taken as 
   $\Delta(l) = c'/l^2$, and it derives from the fact that 
   a finite-size cluster is fully polarized by a lower field 
   than the one of the infinite system. The total 
   conjugate magnetization of each cluster is then taken
   as $\tilde{M}_l(\tilde{h}) = \tilde{h}^{\gamma'} \Theta(\tilde{h})$.
   The above formulas are the same as the ones for the low-field
   Bose-glass phase with $h_{c1}^{(0)} = 0$. We can therefore borrow
   directly the result of Eq.~(\ref{e.expmagn}) and obtain, in the
   high-field Bose glass phase
   \begin{equation}
    S - m_u(h) \sim \exp\left[ -\frac{2|\ln(1-p)|} 
  {c'\left(h_{c2}^{(0)}-h\right)}\right] .
  \label{e.expmagn2}
   \end{equation}

\begin{figure}[h]
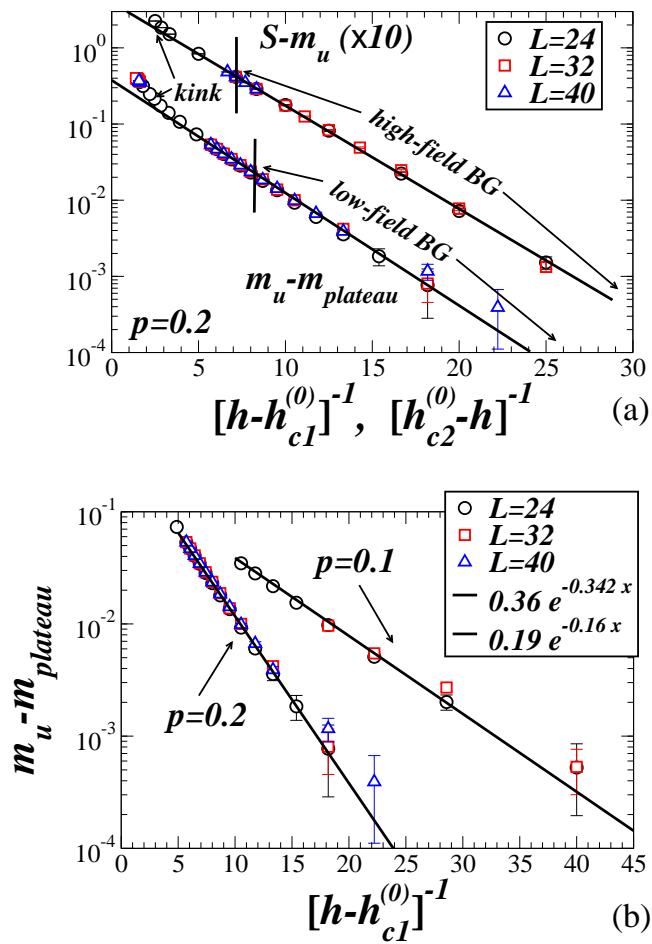

\begin{center}
\null~~~~~~\includegraphics[
     width=78mm,angle=0]{fig7a.eps}
     \vskip .5cm
 \includegraphics[
     width=85mm,angle=0]{fig7b.eps}    
\vskip 1cm
\caption{(a) Exponential behavior of the magnetization curves
in the low- and high-field Bose-glass phase of the site-diluted
bilayer with $p=0.2$. The vertical bars indicate the estimate 
of the transition field from Bose glass to superfluid.
The dashed lines are exponential fits $A\exp(-Bx)$.
The field value at which the magnetization curve exhibits
a kink (see Fig. \ref{f.hscan}) is also indicated.
(b) Same as in the upper panel for the low-field 
Bose-glass phase at two values of site dilution ($p=0.1$ and $p=0.2$).}
\label{f.expmagn}
\end{center}
\end{figure}

 Fig. \ref{f.expmagn} shows the magnetization behavior in the 
two Bose-glass phases of the system. Remarkably the magnetization
curve has a marked exponential dependence on
the inverse of the distance of the field from the clean 
critical fields $h_{c1}^{(0)}$ and $h_{c2}^{(0)}$, reflecting directly 
the rare-event statistics. Also, such exponential dependence
clearly appears to spread well beyond the Bose glass phase
and to hold also in part of the superfluid phase. In particular
we can relate the kink in the magnetization, leading to a
blip in $\chi_u$ as seen in Fig. \ref{f.hscan} 
(see Sec. \ref{sec.results}), to the point
where the two exponential behaviors of the magnetization
coming from $h^{(0)}_{c1}$ and $h^{(0)}_{c2}$ 
[Eq.~(\ref{e.expmagn}) and Eq.~(\ref{e.expmagn2})]
 meet each other roughly halfway at 
 $h \approx (h^{(0)}_{c1} + h^{(0)}_{c2})/2 \approx 1.12$
in a discontinuous manner.

The local-gap
model predictions of Eq.~(\ref{e.expmagn})-(\ref{e.expmagn2})  
is strikingly verified in both Bose-glass phases, as shown in 
Fig. \ref{f.expmagn}(a): the exponential dependence of  
on $\left(h-h_{c1}^{(0)}\right)^{-1}$, 
$\left(h_{c2}^{(0)}-h\right)^{-1}$ is verified over about
three orders of magnitude. In the low-field phase
 we still observe slight deviations from the perfectly
 exponential behavior close to $h_{c1}^{(0)}$, most likely due to the  
 the presence of the free moments. The free moments
 are indeed fully polarized in this phase, with a gap
  to all excitations which grows linearly with
  the field distance from the onset of the plateau phase, 
  $\Delta_{\text{free-moments}} \sim h-h_{\rm plateau}$.
  Yet, for $h \gtrsim h_{c1}^{(0)}$,
  this gap might still not be big enough to completely
  rule out off-resonant exchanges of a triplet
  excitation between the clean regions and the
  free moments, mediated by the $J'$ couplings between
  unpaired spins and intact dimers. When the density
  of triplets in the clean regions is very low,
  namely for $m(h)-m_{\rm plateau} \ll 1$, this 
  small effect might lead to visible deviations with 
  respect to the predictions of the simple local-gap model.
  At high fields, on the other hand, the gap to excitations of
  the free moments is larger, and the above effect 
  is expected to be suppressed, which is consistent 
  with the excellent agreement we find between the
  numerical data and the magnetization of the local-gap 
  model. 
 
  To further test the validity of the local-gap model, 
  we have investigated the low-field Bose-glass phase 
  for a second value of site dilution $p=0.1$, as shown
  in Fig. \ref{f.expmagn}(b) . The 
  exponential nature of the magnetization curve is 
  evident also for this dilution value. Moreover the
  local-gap model predicts the slope of the 
  magnetization curve in logarithmic scale to decrease
  in absolute value with decreasing $p$ as $|\ln(1-p)|$.
  We have tested this result by fitting the data 
  at $p=0.1$ and $p=0.2$ through $A\exp(-Bx)$
  with $x = (h-h_{c1}^{(0)})^{-1}$. According to Eq.~(\ref{e.expmagn})
  we should get
  $B_{p=0.2}/B_{p=0.1} = (\ln 0.8) /(\ln 0.9) = 0.4721... $;
  we numerical obtain $B_{p=0.2}/B_{p=0.1} \approx 0.16/0.342 = 0.468...$,
  in very good agreement with the above prediction. 
  This further demonstrates that the details of the exponential
  magnetization curve are very sensitive to the geometric
  features of the system. This result could in principle 
  be used experimentally, \emph{e.g.}
  to determine the doping concentration in a system, when its
  magnetization curve is compared with a reference system 
  with known doping.

\section{Experimental implications and conclusions}\label{sec.conclu}
  
  In this paper we have presented a complete picture of the 
  extremely rich phase diagram displayed by site-diluted weakly-coupled
  dimer systems as a function of the applied field. We have shown 
  that the field-induced Bose-Einstein condensation of triplet 
  quasiparticles/singlet quasiholes appearing in the clean limit
  is strongly affected by disorder, which introduces 
  a novel Bose-glass phase of quantum localized quasiparticles/quasiholes
  between the insulating (empty) phase and the condensate phase. 
  The quantum phase transition to the ordered phase takes the
  nature of a localization-delocalization transition, and it
  numerically verifies the prediction \cite{Fisheretal89} for 
  the dynamical critical exponent of the Bose-glass to superfluid 
  transition. In the Bose glass phase, we show that rare event 
  statistics dominates the response to the applied magnetic 
  field, and the magnetization curve acquires an unconventional
  exponential dependence on the field reflecting the probability
  distribution of rare clean regions in the sample, as 
  accurately predicted by a local-gap model.
  Finally, in zero field, disorder introduces free moments
  in the dimer-singlet state which show long-range antiferromagnetic
  ordering. We find that this antiferromagnetic state is destroyed
  by the field in an unconventional way, namely \emph{without} full 
  polarization of the free moments. A possible scenario for this
  mechanism involves the formation 
  of finite-range singlets, reminescent of the random-singlet phase
  in bond-disordered $S=1/2$ Heisenberg spin chains \cite{Fisher94}.
  Such singlets survive the application of a small enough field
  which is on the contrary able to fully polarize
  the spins involved in the N\'eel ordered state.
  Further studies are being currently pursued to verify this
  scenario quantitatively \cite{Rongetal06}. 
  
   We believe that our results have immediate experimental
   relevance for all unfrustrated spin gap systems in dimensions $d=2$
   and higher, and with a gap of the order of the strongest magnetic 
   coupling in the system. As already mentioned in the introduction,
   it is generally accepted that the field-induced
   ordering in unfrustrated and magnetically isotropic spin gap 
   systems has the common nature of a Bose-Einstein condensation of 
   quasiparticle excitations \cite{Rice02,Nikunietal00,
   Jaimeetal04,Sebastianetal05,
   Rueggetal03,Affleck91, GiamarchiT99, Matsumotoetal04,Kawashima04,
   MisguichO04,Wesseletal}. The disorder effects 
   on such transition, here discussed in the specific example of the 
   bilayer Heisenberg antiferromagnet, can be fully
   rephrased in the context of coupled Haldane chains,
   coupled spin ladders, and other two- and three-dimensional 
   arrangements of weakly coupled $S=1/2$ dimers. The localization 
   effects leading to the appearence of a Bose-glass phase are 
   possibly more dramatic in quasi one-dimensional systems,
   leading to a larger extent in field values for the 
   Bose-glass phase.
   
    Candidate coupled-dimer compounds for the observation of the phase succession
   of Fig. \ref{f.phasesketch} should display a significant 
   difference between the intradimer coupling $J$ and the overall energy 
   of the interdimer ones, $\sim z\bar{J'}/2$. 
   Here $\bar{J'}$ is the average energy
   of the interdimer couplings and $z$ is the dimer-lattice coordination
   number.
   This requirement leads to a zero field gap $\Delta \approx J -z\bar{J'}/2 \sim J$, 
   which in turn results into a zero field correlation length $\xi_0 \lesssim 1$. 
   If this condition is satisfied in the clean limit, as observed in the 
   bilayer with $g=8$, in the doped case the free moments induced
   in the system are very weakly interacting on average, 
   $\langle J_{\rm eff}\rangle /J \sim p (z\bar{J'}/J)^2 \exp(-1/\xi_0) \ll 1$, 
   and their 
   zero-field ordered state is destroyed by an applied field
   much smaller than the clean-limit gap $\Delta$. This 
   guarantees a large separation of energy scales between
   the order-by-disorder phenomenon and the field-induced
   condensation of triplet quasi-particles, hence offering
   the possibility of clearly observing the intermediate phases 
   (disordered-free-moment, plateau phase and Bose-glass).
   If the above requirement is not satisfied by the intrinsic
   parameters $J$, $\bar{J'}$ and $z$ of the system, there is still the
   possibility of working at very low doping $p \ll 1$.
   Fortunately, the requirement $J\ll z\bar{J'}/2$ appears to be met 
   by a variety of compounds. Among others: BaCuSi$_2$O$_6$, 
   which has  $z\bar{J'}/2J \approx 0.3 $~ \cite{Sebastianetal05},
   Sr$_2$Cu(BO$_3$)$_2$ which has $z\bar{J'}/2J \sim 0.2-0.3$
   ~ \cite{Sebastianetal05-2}, Cu$_{12}$(C$_5$H$_{12}$N$_2$)$_2$Cl$_4$
   with $z\bar{J'}/2J \sim 0.2$ ~\cite{Chaboussantetal97},
   (C$_5$H$_{12}$N$_2$)$_2$Cu Br$_4$ with $z\bar{J'}/2J \sim 0.3$
   ~\cite{Watsonetal01}, etc. 
    On the other hand, if the inter-moment coupling 
   is too large, \emph{short-range} antiferromagnetic
   correlations between the free moments might survive up to 
   a field of the order of the gap, thus eliminating the plateau 
   phase, as theoretically observed in the bilayer system with 
   $g=4$ ~\cite{RoscildeH05}. If \emph{long-range} antiferromagnetic
   order of the free moments persists up to fields of the
   order of the gap, the order-by-disorder phase might even merge
   with the ordered phase of the intact dimers, so that all
   disordered phases in between are washed out \cite{Mikeskaetal04}. 
   This is the conclusion of a recent study on the Mg-doped
   TlCuCl$_3$,
   which has indeed $z\bar{J'}/2J \sim 0.9$ ~\cite{preprint06}.
   
   In the case of coupled Haldane chains with non-magnetic doping, 
   $S=1/2$ moments are induced by impurities at the edges of finite chain 
   segments \cite{MiyashitaY93}. To have a large energy separation
   between the inter-moment interaction and the Haldane gap, it is
   certainly necessary that the gap $\Delta \approx 0.4 J$ ~\cite{WhiteH93}
   be much larger than the characteristic energy of the interchain 
   couplings, $z\bar{J'}\xi_0$,  where $\bar{J'}$ is the average interchain 
   coupling, $z$ is the coordination number of the coupled chains, and 
   $\xi_0\sim 6$ ~\cite{WhiteH93} is the correlation length of the isolated chain. 
   Nonetheless, for weakly coupled chains the dominant coupling
   channel between the free moments is along the chain direction, 
   namely it depends crucially on the average spacing between
   impurities along each chain $\sim 1/p$ compared with the 
   characteristic decay length of the effective interactions,
   which is given by $\xi_0$ for small impurity 
   concentrations. Therefore it is necessary that $1/p \gg \xi_0$,
   which roughly means $p \ll 0.1$. All these requirements have
   evidently been met by recent experiments \cite{Uchiyamaetal99} 
   on Pb(Ni$_{1-p}$Mg$_p$)$_2$V$_2$O$_8$ with $p \leq 0.02$.
   To our knowledge, this system stands as the only experimental 
   example in which first the impurity-induced ordered phase is destroyed 
   by the field through polarization of the free moments, and then order 
   is induced again by the field through partial polarization of the clean 
   chain segments. The temperature at which the experiments have been 
   performed so far appears to be too high to resolve the predicted 
   succession of disordered phases in between the two ordered
   ones, but further experiments at lower temperatures are a very 
   promising test of the scenario presented in this paper.

 \section{Acknowledgements}
  We acknowledge fruitful discussions and correspondence with 
  F. Alet, C. Lhuillier, R. Yu and M. Vojta. A special thanks
  goes to S. Haas for his help, advise and support throughout 
  this project. 
  This work is supported by DOE, and by the European Union 
  through the SCALA integrated project. Computational 
  facilities have been generously provided by the HPC Center at USC,
  and by NERSC. 

\appendix
\section{Wavefunction and effective interaction of the free moments in 2$d$}\label{app.wf}
In this appendix we perturbatively derive the wavefunction 
of the free moments, Eq.~(\ref{e.wf}),
and their effective interaction in two dimensions, Eq.~(\ref{e.jeff}). 
We first consider a \emph{single} unpaired spin at site $i$ surrounded by 
intact dimers, and calculate the first-order perturbative
correction to the ground state due to the coupling
between the unpaired spin and the host matrix,  
$V_i = J' \sum_{\delta} \bm{S}_i\cdot \bm{S}_{i+\delta}$, where $\delta$ connects
the site $i$ to its nearest neighbors.  The unperturbed state
of the system for $J'=0$ is assumed as 
$|\Psi^{(0)}\rangle = |\uparrow_i\rangle|0\rangle$, where $|\uparrow_i\rangle$ is
an arbitrary state of the unpaired spin and
$|0\rangle$ is the
unperturbed dimer-singlet state of the surrounding
intact dimers. The perturbation term
\begin{equation}
V^{-+}_i = \frac{J'}{\sqrt{N}} 
\sum_{\bm q} z \gamma_{\bm q}~ S_i^{-} S_{\bm q}^{+}
~e^{i{\bm q}\cdot {\bm r}_i}~~~~
\left(\gamma_{\bm q} = 1/z \sum_{\delta} e^{i {\bm q}\cdot {\bm \delta}}\right)
\end{equation}   
flips the unpaired spin and transfers its magnetic moment
to a triplet excitation of the intact dimers, namely it 
mixes up the unperturbed ground state $|\Psi^{(0)}\rangle$
with excited states of the form
\begin{equation}
|\Psi({\bm k})\rangle = |\downarrow_i\rangle |{\bm k}\rangle.
\end{equation}
where $|{\bm k}\rangle$ is the state of the intact
dimers with one elementary triplet excitation at
momentum $\bm k$.
 Assuming the following dispersion
relation for the long-wavelength gapped modes of the
unperturbed coupled dimer system
\begin{equation}
\epsilon({\bm k}) = \sqrt{\Delta^2 + v^2 k^2}
\end{equation}
we obtain the first-order perturbed state
\begin{eqnarray}
 |\Psi^{(1)}\rangle &=& |\Psi^{(0)}\rangle + \frac{J'z}{\sqrt N}
 \sum_{\bm k} 
 \frac{{\cal M}_{\bm k} e^{i{\bm k}\cdot {\bm r}_i}}{\epsilon({\bm k})}
 |\downarrow_i\rangle |{\bm k}\rangle \nonumber \\
 &=& |\Psi^{(0)}\rangle + 
 |\downarrow_i\rangle \sum_{\bm r} \psi(\bm r)  |\bm r\rangle
 \label{e.sumk}
 \end{eqnarray}
with ${\cal M}_{\bm k} \approx \gamma_{\bm k}
\langle \bm k | S^{+}_{\bm k} |0\rangle$.
In the last step we have introduced the localized
triplet states 
$|\bm r\rangle = 1/\sqrt{N} \sum_{\bm q} 
\exp(-i\bm q\cdot\bm r) |\bm q\rangle$. In this form 
the amplitude $\psi(\bm r)$ takes the meaning of
\emph{wavefunction} of the free moment which is 
transferred by the perturbation from the unpaired
spin to the host dimer system. 
Passing to the continuum limit, and neglecting
the ${\bm k}$-dependence of the matrix elements
${\cal M}_{\bm k}\approx 1$, we obtain
\begin{equation}
\psi(\bm r) = J'z \int d^2 k ~
\frac{e^{i{\bm k}\cdot {\bm r}}}{\epsilon(\bm k)} =
\frac{J'z}{\Delta \xi_0} \int_0^{\infty} 
\frac{k J_0(kr)}{\sqrt{\xi_0^{-2}+k^2}}
\label{e.step1}
\end{equation}
where $\xi_0 = v/\Delta$ is the correlation length of
the unpertubed host dimers and $J_0(x)$ is an ordinary 
Bessel function. The last expression contains an 
Hankel-Nicholson-type integral \cite{AbramowitzS65}
whose solution is expressed through the modified
Bessel function $K_{1/2}$ in the form
 \begin{equation}
\psi(\bm r) =  \frac{J'z}{\Delta \xi_0} \sqrt{\frac{2}{\pi}}
~\frac{K_{1/2}(r/\xi_0)}{\sqrt{r/\xi_0}}.
\label{e.step2} 
\end{equation}
The asymptotic expression \cite{AbramowitzS65} of
Eq.~(\ref{e.step2})
for large $r$ gives Eq.~(\ref{e.wf}).

The effective interaction between the free moments arises instead
from second order perturbation corrections in the ground-state
energy of the system \cite{SigristF96,Mikeskaetal04}. 
Taking two unpaired spins at sites $i$ and $j$, they
exchange a triplet excitation through second-order
perturbations 
$V_i^{-+} V_j^{+-}$ and $V_i^{+-} V_j^{-+}$. These
terms introduce a singlet-triplet splitting for the
joint state of the unpaired spins which 
immediately gives the effective interaction strength
$J_{\rm eff}(i,j;\alpha,\beta)$ in the form
\begin{equation}
J_{\rm eff}(i,j;\alpha,\beta) = (-1)^{\sigma}
\frac{(J'z)^2}{N} \sum_{\bm k} |{\cal M}_{\bm k}|^2
\frac{e^{i{\bm k}\cdot ({\bm r_i}-{\bm r_j})}}
{8\epsilon(\bm k)}
\end{equation}
where $\sigma = x_i + y_i + x_j + y_j +\alpha+\beta$.
We then recover a similar expression to that
encountered in Eq.~(\ref{e.sumk}), which, in the
continuum limit, can be evaluated through 
ordinary and modified Bessel functions as in
Eq.~(\ref{e.step1})-(\ref{e.step2}). 
The final expression is
\begin{equation}
J_{\rm eff}(i,j;\alpha,\beta) = (-1)^{\sigma} 
\frac{(J'z)^2}{4\pi\Delta\xi_0} \frac{\sqrt{2}}{\Gamma(1/2)}
\frac{K_{1/2}(r/\xi_0)}{\sqrt{r/\xi_0}}
\end{equation}
where $r = |{\bm r_i}-{\bm r_j}|$. In the large-$r$ limit
we then recover Eq.~(\ref{e.jeff}).

\end{document}